\documentclass[aps,pre,twocolumn,groupedaddress,showpacs]{revtex4-1}
\usepackage{graphicx}
\usepackage{dcolumn}
\usepackage{bm}
\usepackage{epsfig}
\usepackage{enumerate}

\usepackage{color}
\definecolor{dblue}{rgb}{0,0,0.75}
\definecolor{dred}{rgb}{0.6,0,0}
\definecolor{dgreen}{rgb}{0,0.5,0}

\input epsf
\epsfclipon

\begin{document}
\title{Phase transitions in the $q$-voter model with noise on a duplex clique}
\author{Anna Chmiel, Katarzyna Sznajd-Weron} 
\affiliation{Department of Theoretical Physics, Wroclaw University of Technology, Wroclaw, Poland}
\date{\today}
\begin{abstract}
We study a nonlinear $q$-voter model with stochastic noise, interpreted in the social context as independence, on a duplex network. To study the role of the multi-levelness we propose three methods of transferring the model from a mono- to a multiplex network. They take into account two criteria -- one related to the status of independence (\texttt{LOCAL} vs. \texttt{GLOBAL}) and one related to peer pressure (\texttt{AND} vs. \texttt{OR}). 
In order to examine the influence of the presence of more than one level in the social network, we perform simulations on a particularly simple multiplex --  a duplex clique, which consists of two fully overlapped complete graphs (cliques). Solving numerically the rate equation and simultaneously conducting Monte Carlo simulations, we provide evidence that even a simple rearrangement into a duplex topology may lead to significant changes in the observed behavior. However, qualitative changes in the phase transitions can be observed for only one of the considered rules -- \texttt{LOCAL\&AND}. For this rule the phase transition becomes discontinuous for $q=5$, whereas for a monoplex such a behavior is observed for $q=6$. Interestingly, only this rule admits construction of realistic variants of the model, in line with recent social experiments.

\end{abstract}
 \maketitle

\section{Introduction}
Opinion dynamics is one of the most investigated subfields of sociophysics \cite{Hol:Kac:Sch:01,Cas:For:Lor:09,Gal:12,Sen:Cha:13}. Subjectively, there are at least two important reasons why physicists study this topic. The first motivation comes from social sciences and can be described as a temptation to build a bridge between the micro and macro levels in describing social systems. Traditionally, there are two main disciplines that study social behavior - sociology and social psychology. Although the subject of the study is the same for both disciplines, the usually taken approach is very different. Sociologists study social systems from the level of the social group whereas social psychologists concentrate on the level of the individual \cite{Mye:13}. From the physicist's point of view this is similar to the relationship between thermodynamics and statistical physics. This analogy raises the challenge to describe and understand the collective behavior of social systems (sociology) from the level of interpersonal interactions (social psychology). The second motivation to deal with opinion dynamics is related to the development of non-equilibrium statistical physics. Models of opinion dynamics are often very interesting from the theoretical point of view \cite{Kra:Red:Ben:10}. One of the best examples is the famous voter model or, the recently introduced, nonlinear $q$-voter model \cite{Cas:Mun:Pas:09}. Both models are based on dichotomous opinions and belong to a wide class of binary-state dynamics \cite{Hol:Kac:Sch:01,Gal:05,Gle:13,Roy:Bis:Sen:14}. It should be stressed here, that binary opinions are natural from the social point of view, since dichotomous response format with 1 (yes, true, agree) and 0 (no, false, disagree) as response options is one of the most common in social experiments \cite{Byr:Kai:12,Rob:Fra:Kue:07}. 

Among many other binary opinion models \cite{Cas:For:Lor:09,Gal:12,Sen:Cha:13}, the $q$-voter model is not only interesting from theoretical point of view, but also justified from the social point of view. In short, within the $q$-voter model each individual interacts with a set of $q$ neighbors (a $q$-lobby) and if all $q$ neighbors share the same state (i.e. the $q$-lobby is unanimous), the individual conforms to this state. As originally proposed, in the other case (disagreement) the individual changes its state with probability $\epsilon$ \cite{Cas:Mun:Pas:09}. However, in some later publications the model with $\epsilon=0$ was studied, as a natural generalization of the Sznajd model \cite{Prz:Szn:Tab:11,Nyc:Szn:Cis:12,Tim:Pra:14}. The unanimity rule can be justified based on social experiments. It has been observed in number of experiments that a small unanimous group may be more efficient than a much larger group with a non-unanimous majority \cite{Mye:13}. In a classical series of experiments on conformity, Solomon Asch has found that the presence of a social supporter reduced conformity dramatically --  participants of the experiment were far more independent when they were opposed by a seven person majority and had a partner sharing the same opinion than when they were opposed by a three-person majority and did not have a partner \cite{Asch:55}. Influence of a consistent minority on the responses of a majority has been reported by Moscovici et al. \cite{Mos:Lag:Naf:69}. Also recent neurological experiments suggest that unanimous opinions may be critical for normative influence \cite{Cam:Bac:Roe:Dol:Fri:10}.

From the physicist's point of view the $q$-voter model is interesting because of the rich behavior related to phase transitions 
\cite{Cas:Mun:Pas:09,Mor:Liu:Cas:Pas:13,Nyc:Szn:Cis:12} as well as the controversy related to the exit probability of the model \cite{Prz:Szn:Tab:11,Tim:Pra:14}. In this paper we will focus on phase transitions driven by stochastic noise which -- in the social context -- may be interpreted as independence \cite{Nyc:Szn:Cis:12,Nyc:Szn:13}. In social psychology, independence is recognized as one of the two types of nonconformity and means resisting influence \cite{Nai:Mac:Lev:00}. It has been noticed that independence plays a role similar to the temperature and introduces order-disorder phase transitions \cite{Nyc:Szn:13}. Interestingly, it has been shown that in the case of a complete graph, the phase transition changes its type from continuous to discontinuous for $q \ge 6$ \cite{Nyc:Szn:Cis:12,Nyc:Szn:13}.

Up till now, the $q$-voter model has been studied on monoplex networks, i.e. networks that consist of only one level. However, as noted recently, interactions among individuals can be of qualitatively different nature and therefore modeled by multi-level networks \cite{przeg_bianc}. In the last two years, a lot of attention has been devoted to the analysis of various dynamics on multiplex networks, including diffusion processes \cite{dyf1}, epidemic spreading \cite{epi_Arenas,epi_plos,prx_moreno} and voter dynamics \cite{marina}.  
Brummitt et al. \cite{Goh1} have generalized the threshold cascade model on complex networks \cite{Watt}. In \cite{Goh2} they further expanded the model and introduced the idea of OR and AND nodes. An OR node is activated as soon as a sufficiently large fraction of its neighbors are active in at least one level. An AND node is activated only if in each and every layer a sufficiently large fraction of its neighbors are active. We will use a related notion of OR and AND types of social influence. 

Without doubt most real-world social networks consist of many levels. For instance, a student may belong to a network of classmates, a network of sport-club teammates and a network of Facebook friends. The question is if this multi-levelness is important for the macroscopic (or global) properties of the social system, such as public opinion, or not. We try to answer this question within the $q$-voter model that takes into account two types of response to social influence -- conformity and independence.

The remainder of this paper is organized as follows. In Sec. II, we briefly recall the $q$-voter model on a single monoplex clique. We extend this framework to multiplex networks in Sec. III and propose three rules (\texttt{GLOBAL\&AND}, \texttt{GLOBAL\&OR} and \texttt{LOCAL\&AND}). In Sec. IV we derive rate equations, that describe the time evolution of the system, for each of the three rules. Furthermore in Sec. V, based on these equations, we derive phase diagrams and compare results obtained from analytical equations with those obtained from Monte Carlo simulations. The next section is devoted to a deeper analysis of the phase transitions. We wrap up the results and conclude in Sec. VII.

\section{The $q$-voter model on a single monoplex clique}

The $q$-voter model with the stochastic noise, interpreted in the social context as independence, has been already analyzed on a monoplex complete graph \cite{Nyc:Szn:Cis:12}.  Here, following \cite{Nyc:Szn:Cis:12}, we also consider a set of $N$ individuals, which are described by the binary variables $S_i=\pm 1$ (spins 'up' or 'down'). At each elementary time step $t$ we randomly choose an $i$-th node (i.e. a voter) and a $q$-lobby, which is a randomly picked group of $q$ individuals. Only if the $q$-lobby is self-consistent it can  influence the voter. With probability $1-p$ the $q$-lobby (if it is homogeneous) acts on the state of the voter, which means that the voter changes state to the state of the $q$-lobby. With probability $p$ voter behaves independently -- with equal probabilities flips to the opposite direction $S_i(t+1)=-S_i(t)$ or keeps its original state $S_i(t+1)=S_i(t)$. Therefore only the following changes are possible: 
\begin{eqnarray}
\underbrace{\uparrow\uparrow \ldots \uparrow}_{q} \Downarrow & \stackrel{1-p}{\longrightarrow} & \underbrace{\uparrow\uparrow \ldots \uparrow}_{q} \Uparrow,\nonumber \\
\underbrace{\downarrow\downarrow \ldots \downarrow}_{q}\Uparrow & \stackrel{1-p}{\longrightarrow} & \underbrace{\downarrow\downarrow \ldots \downarrow}_{q}\Downarrow, \nonumber \\
\underbrace{\ldots}_{q} \Downarrow & \stackrel{p/2}{\longrightarrow} & \underbrace{\ldots}_{q} \Uparrow, \nonumber \\
\underbrace{\ldots}_{q} \Uparrow & \stackrel{p/2}{\longrightarrow} & \underbrace{\ldots}_{q} \Downarrow, 
\label{mono}
\end{eqnarray}

where a single-line arrow represents the state of a node belonging to the $q$-lobby while the state of the voter is marked with a double-line arrow. It has been shown that the system, described by the $q$-voter model with independence, undergoes the phase transition at $p=p_c(q)$. For $p<p_c$ the majority coexists with the minority opinion (ordered state) and for $p>p_c$ there is a status-quo (disordered state) \cite{Nyc:Szn:Cis:12}. Interestingly, it occurred that for $q \le 5$ the phase transition is continuous, whereas for $q>5$ becomes discontinuous. 

The most natural quantity that describes the macroscopic behavior of such a system is magnetization, which from the social point of view represents so called public opinion:
\begin{equation}
m(t)=\frac{1}{N} \sum_{i=1}^N S_i(t).
\label{eq_mag}
\end{equation}
Moreover, in the case of a complete graph, the magnetization fully describes the state of system. In this paper we will calculate it in two ways -- by Monte Carlo simulations of the microscopic system of the size $N$ and by numerical solution of the equation describing the time evolution of the average magnetization. In the case of the Monte Carlo simulations we will calculate ensemble average of the magnetization in the stationary state:
\begin{equation}
<m>=\frac{1}{M} \sum_{j=1}^M m_i,
\label{eq_mag1}
\end{equation}
where $m_i$ denotes the stationary value of the magnetization in $i$-th realization (sample) and $i=1,\ldots,M$. In this paper we average all Monte Carlo results over $M=10^3$ samples.

\section{The $q$-voter model on a duplex clique}

\begin{figure}
\vskip 0.3cm
\centerline{\epsfig{file=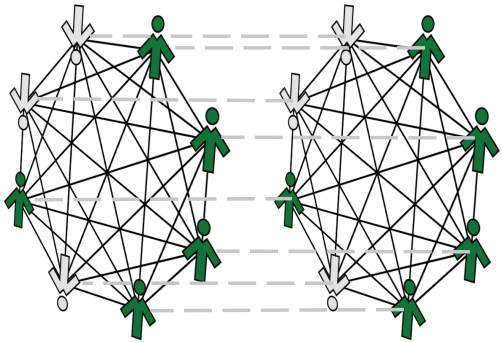,width=.8\columnwidth}}
\caption{The topology of a duplex clique, i.e. a network that consists of two distinct levels (layers), each of which is represented by a complete graph (i.e. a clique) of size $N$. Levels represent two different communities but are composed of exactly the same people -- each node possesses a counterpart node in the second level. Interconnections between levels (denoted by the gray dashed lines) are realized exclusively by connecting the node with its own counterpart (i.e. the same node) on the other level. Links within a single level (black solid lines) represent some kind of social relation (e.g. friendship). }
\label{top_tak}
\end{figure}

Let us start with defining a duplex clique, which is a particular case of a multiplex. Specific definitions of multiplex networks have been introduced in \cite{przeg_bianc,przeg_arenas,tenor}. Such systems consist of distinct levels (layers) and the interconnections between levels are only between a node and its counterpart in the other layer (i.e. the same node). Here we consider a duplex clique (see Fig. \ref{top_tak}), i.e. a network that consists of two distinct levels (layers), each of which is represented by a complete graph (i.e. a clique) of size $N$. Levels represent two different communities (e.g. Facebook and school class), but are composed of exactly the same people -- each node possesses a counterpart node in the second level. Such an assumption reflects the fact that we consider fully overlapping levels, being an idealistic scenario. We also assume that each node possesses the same state on each level, which means that the society consists of non-hypocritical individuals only.

\begin{figure}
\vskip 0.3cm
\centerline{\epsfig{file=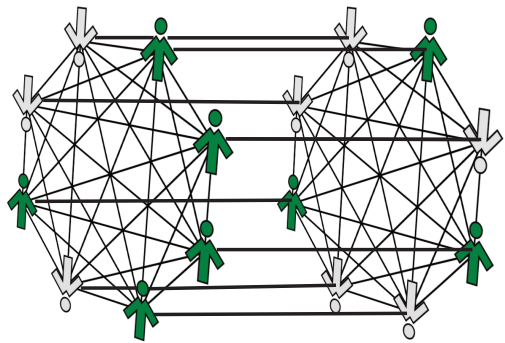,width=.8\columnwidth}}
\caption{The topology of two inter-connected monoplex cliques. This network consists of two levels, each of which being represented by a complete graph of size $N$. Levels represent two different communities and are composed by different people. Links within a single level are equal to the connections between levels (black solid lines) and represent some kind of social relation (e.g. friendship).}
\label{top_nie}
\end{figure}

It is worth to stress the difference between a duplex clique and two inter-connected monoplex cliques (see Fig. \ref{top_nie}), where the state of a node on one level is not directly related to the state of a node in the second layer. In inter-connected monoplex cliques the inter-clique links fulfill the same role as intra-clique edges. Classical voter models were analyzed on inter-connected cliques in \cite{voter_two,Redner}.

In this paper we investigate the $q$-voter model on a duplex clique. As in the case of the monoplex, we consider a set of $N$ individuals described by the binary variables $S_i=\pm 1$, which are the same on both levels. On each level all individuals are connected with each other and therefore create a duplex clique, as shown in Fig. \ref{top_tak}.

\begin{table}
\setlength{\tabcolsep}{4.5pt}
\begin{center}
\begin{tabular}{|l|c|c|}
\hline  & \texttt{AND} & \texttt{OR} \\
\hline
global independence & (i) \texttt{GLOBAL\&AND}  & (ii) \texttt{GLOBAL\&OR} \\\hline
local independence & (iii) \texttt{LOCAL\&AND}   & x \\\hline
\end{tabular}
\end{center}
\caption{Three versions of dynamics on the multiplex clique. We do not consider the \texttt{LOCAL\&OR} rule because of the difficulty of such a concept related to social unreality and algorithmic ambiguity.} \label{tab1}
\end{table}

We consider two criteria of level dependence -- one related to the status of independence (GLOBAL vs. LOCAL) and one related to peer pressure (AND vs. OR). (see also Tab. \ref{tab1}):
\begin{enumerate}
\item Criteria related to the status of independence: the \texttt{GLOBAL} rule means that an agent is independent on both levels, but the \texttt{LOCAL} rule admits a situation where a person is independent in one clique but not in the other.
\item Criteria related to the peer pressure: the \texttt{AND} dynamics is more restrictive and a node changes its state only if {\it both levels} suggest changes, in the \texttt{OR} variant {\it one level} is enough to change the state of an individual.
\end{enumerate}

Finally we propose the following three rules:
\begin{enumerate}[(i)]
\item \texttt{GLOBAL\&AND} -- global independence and the \texttt{AND} rule (see an example in Fig. \ref{and_global_rule})\\
With probability $p$ the voter is independent and with $1-p$ behaves like a conformist regardless of the level. In the case of independence the voter changes its state to the opposite one with probability $1/2$ (we automatically change the state of the voter on both levels). In the case of conformity the voter changes its state only when both $q$-lobbies (i.e. on the first and on the second level) are homogeneous and both have the state opposite to the state of the voter. Therefore only the following changes are possible: 
\begin{eqnarray} \label{AND_glob}
\frac{\downarrow \downarrow \ldots \downarrow}{\downarrow \downarrow \ldots \downarrow}  \Uparrow & \stackrel{1-p}\longrightarrow & \frac{\downarrow \downarrow \ldots \downarrow}{\downarrow \downarrow \ldots \downarrow}  \Downarrow  \nonumber  \\
\frac{\uparrow \uparrow \ldots \uparrow}{\uparrow \uparrow \ldots \uparrow} \Downarrow & \stackrel{1-p}\longrightarrow & \frac{\uparrow \uparrow \ldots \uparrow}{\uparrow \uparrow \ldots \uparrow} \Uparrow  \nonumber  \\
\frac{\ldots}{\ldots}  \Downarrow & \stackrel{p/2}\longrightarrow & \frac{\ldots}{\ldots} \Uparrow  \nonumber  \\
\frac{\ldots}{\ldots}  \Uparrow & \stackrel{p/2}\longrightarrow & \frac{\ldots}{\ldots} \Downarrow,
\end{eqnarray}
where a single-line arrow represents the state of a node belonging to the $q$-lobbies: arrows in the numerator represent states of the nodes belonging to the $q$-lobby chosen on first level and arrows in the denominator those on second level; voter's state is marked by a double-line arrow. 

We should commend here, that the $q$-voter model with the \texttt{GLOBAL\&AND} rule and $q=q_2$ on the duplex clique is equivalent to the $q$-voter on the monoplex clique with $q=q_1=2q_2$ (i.e. a monoplex clique with a $q$-lobby size twice as large as in the duplex case). This is visible in Fig. \ref{and_global} where we compare the Monte Carlo simulations obtained for the duplex with those obtained for the monoplex topology. 

\begin{figure}
\vskip 0.3cm
 \centerline{\epsfig{file=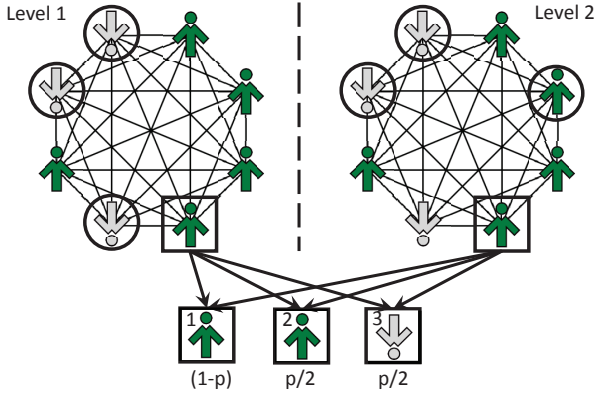,width=.95\columnwidth}}
    \caption{\textbf{The \texttt{GLOBAL\&AND} rule}: a voter is independent regardless of the level with probability $p$ and is subjected to the peer pressure with probability $1-p$ only if $q$-panels on both levels are self-consistent. In this example on level $1$ the $q$-lobby (agents in circles) is homogeneous but on level $2$ the lobby is not self-consistent. Therefore the voter will not change its state under the peer pressure.}
\label{and_global_rule}
\end{figure}

\begin{figure}
\vskip 0.3cm
 \centerline{\epsfig{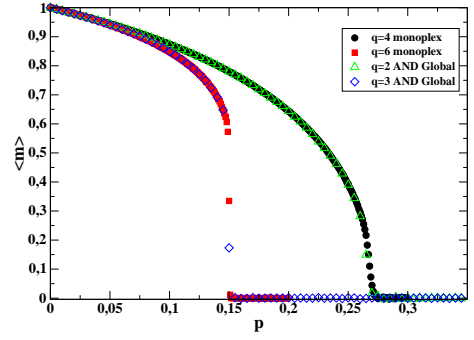}}
    \caption{A comparison between the $q$-voter model on the monoplex clique (full symbols) and the $q$-voter model with the \texttt{GLOBAL\&AND} rule on the duplex clique (empty symbols) -- the \texttt{GLOBAL\&AND} rule leads to a trivial result, identical with the monoplex case for a doubled value of $q$. An ensemble average $<m>$ of the magnetization, as a function of the stochastic noise $p$, was obtained by the Monte Carlo simulations for the system of size $N=10^4$.}
 \label{and_global}
\end{figure}

\item \texttt{GLOBAL\&OR} -- global independence and \texttt{OR} rule (see an example in Fig. \ref{or_global_rule} ).\\
Here, similarly as in the \texttt{GLOBAL\&AND} rule, the status of independence is the same for both levels. With probability $p$ the voter is independent and changes its state to the opposite one with probability $1/2$. With probability $1-p$ the voter behaves like a conformist and its state is dependent on both $q$-lobbies. In contrast to \texttt{AND} dynamics, now the voter changes its state to the opposite one even when only one $q$-lobby is self-consistent and the second is not. In the situation when two $q$-lobbies are homogeneous but not in agreement, i.e. one $q$-lobby supports the voter and second suggests to change its state, the voter becomes confused and stays in its old state:

\begin{equation}
\frac{\uparrow \uparrow \ldots \uparrow}{\downarrow \downarrow \ldots \downarrow}  \Uparrow \stackrel{confuse}\longrightarrow \frac{\uparrow \uparrow \ldots \uparrow}{\downarrow \downarrow \ldots \downarrow} \Uparrow.  
\end{equation}

All situations that lead to change are shown below: 
\begin{eqnarray}
\frac{\downarrow \downarrow \ldots \downarrow}{\downarrow \downarrow \ldots \downarrow}  \Uparrow & \stackrel{1-p}\longrightarrow & \frac{\downarrow \downarrow \ldots \downarrow}{\downarrow \downarrow \ldots \downarrow} \Downarrow,  \nonumber  \\
\frac{\uparrow \uparrow \ldots \uparrow}{\uparrow \uparrow \ldots \uparrow}  \Downarrow & \stackrel{1-p}\longrightarrow & \frac{\uparrow \uparrow \ldots \uparrow}{\uparrow \uparrow \ldots \uparrow} \Uparrow,  \nonumber  \\
\frac{\ldots \downarrow \uparrow \downarrow \uparrow \ldots}{\downarrow \downarrow \ldots \downarrow}  \Uparrow & \stackrel{1-p}\longrightarrow & \frac{\ldots \downarrow \uparrow \downarrow \uparrow \ldots}{\downarrow \downarrow \ldots \downarrow}  \Downarrow,  \nonumber  \\
\frac{\ldots \downarrow \uparrow \downarrow \uparrow \ldots}{\uparrow \uparrow \ldots \uparrow}  \Downarrow & \stackrel{1-p}\longrightarrow & \frac{\ldots \downarrow \uparrow \downarrow \uparrow \ldots}{\uparrow \uparrow \ldots \uparrow}  \Uparrow,  \nonumber  \\
\frac{\uparrow \uparrow \ldots \uparrow}{\ldots \downarrow \uparrow \downarrow \uparrow \ldots}  \Downarrow & \stackrel{1-p}\longrightarrow & \frac{\uparrow \uparrow \ldots \uparrow}{\ldots \downarrow \uparrow \downarrow \uparrow \ldots} \Uparrow, \nonumber \\
\frac{\downarrow \downarrow \ldots \downarrow}{\ldots \downarrow \uparrow \downarrow \uparrow \ldots}  \Uparrow & \stackrel{1-p}\longrightarrow & \frac{\downarrow \downarrow \ldots \downarrow}{\ldots \downarrow \uparrow \downarrow \uparrow \ldots} \Downarrow, \nonumber \\
\frac{\ldots}{\ldots}  \Downarrow & \stackrel{p/2}\longrightarrow & \frac{\ldots}{\ldots} \Uparrow,  \nonumber \\
\frac{\ldots}{\ldots}  \Uparrow & \stackrel{p/2}\longrightarrow & \frac{\ldots}{\ldots} \Downarrow.
\end{eqnarray}

\begin{figure}
\vskip 0.3cm
 \centerline{\epsfig{file=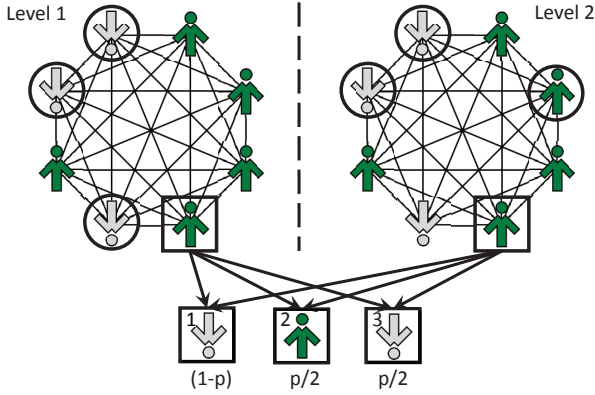,width=.95\columnwidth}}
  \caption{\textbf{The \texttt{GLOBAL\&OR} rule}: a voter is independent regardless of the level with probability $p$ and is subjected to the peer pressure with probability $1-p$ if at least on one level the $q$-panel is self-consistent. In this example on level $1$ the $q$-lobby (agents in circles) is homogeneous and has the state opposite to the state of the voter (an agent in the square). Simultaneously, on level $2$ the $q$-lobby is not self-consistent. Therefore the voter is not confused by two opposite $q$-lobbies and is influenced by the first $q$-lobby.}
\label{or_global_rule}  
\end{figure}

\item \texttt{LOCAL\&AND} -- local independence and \texttt{AND} rule (see an example in Fig. \ref{and_local_rule} ).\\
In this case the independence is related to the level,i.e. we run dynamics separately on each level. It means that a voter is independent on the first level with probability $p$ and with probability $1-p$  behaves as a conformist --- it is under the influence of the $q$-lobby on this level. The same situation is on the second level, where regardless of the first level we choose if the voter behaves independently or conform the $q$-lobby on the second level. Finally we change the state of the voter only when both separated dynamics give in result the same state. 

\begin{figure}
\vskip 0.3cm
 \centerline{\epsfig{file=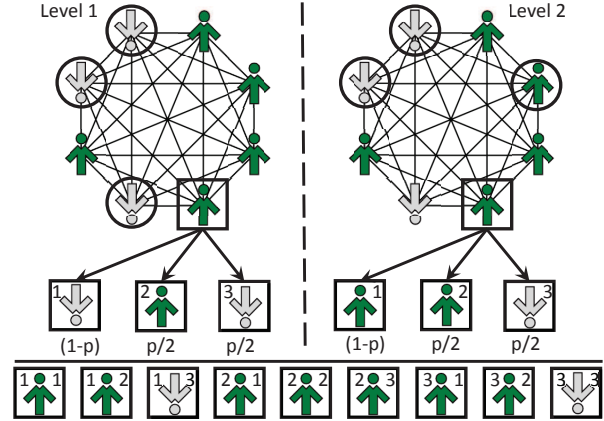,width=.95\columnwidth}}
  \caption{\textbf{The \texttt{LOCAL\&AND} rule}: a voter is independent separately on each level with probability $p$ and subjected to peer pressure separately on each level with probability $1-p$. It means that on each level independently the voter (in the square) can be in one of three possible states: $+1$ with probability $p/2$, $-1$ with probability $p/2$ and in a state suggested by the $q$-lobby (in circles) with probability $1-p$. In this example,  the $q$-lobby is homogeneous on the first level and therefore it influences the voter. On level $2$ the $q$-lobby is not self-consistent and therefore there is no peer pressure. Finally, there are nine possible pairs that represent states on the first and on the second level -- see the numbers in the corners of squares in the bottom line. The voter changes its state only if states, obtained independently on each level, are the same -- see all possible final states of the voter in the bottom line.}
\label{and_local_rule}
\end{figure}
\end{enumerate}
We do not consider the \texttt{LOCAL\&OR} rule because of the difficulty of such a concept related to social unreality and algorithmic ambiguity.

\section{The time evolution}
The aim of this section is to derive equations that describe the time evolution of the system for each of three considered rules. Let us denote by $N_\uparrow(t)$ the number of voters in $+1$-state (up-spins) at time $t$ and by $N_\downarrow(t)$ the number of voters in $-1$-state (down-spins). The total number $N$ of spins in a system does not change and we can define the concentration of up-spins at time $t$ as:
\begin{equation}
c(t)=\frac{N_\uparrow(t)}{N}.
\end{equation} 

Since all individuals keep the same state on both levels and we consider a duplex clique, we can simplify our analysis by considering concentration $c(t)$ only on one level. However, we need to stress that the changes of the state of the node occur under the influence of both levels.
In a single time step $\Delta_t$ three scenarios are possible --- the number of up-spins $N_{\uparrow}(t)$ will either: increase by 1, decrease by 1 or remain constant. Simultaneously the concentration $c(t)$ increases or decreases by $\frac{1}{N}$ or remains constant:

\begin{eqnarray}
\gamma^+(c) & = & \Pr \{c(t+\Delta_t) = c(t)+ \frac{1}{N}\},\\
\gamma^-(c) & = & \Pr \{c(t+\Delta_t) = c(t)- \frac{1}{N}\}, \nonumber \\
\gamma^0(c) & = & \Pr \{c(t+\Delta_t) = c(t) \}=1-\gamma^+(c)-\gamma^-(c) \nonumber.
\label{eq:gamma}
\end{eqnarray}

The time evolution of the average concentration is given by the rate equation:
\begin{equation}\label{master}
<c(t+\Delta_t)>=<c(t)>+\frac{1}{N} \left[ \gamma^+(c) -\gamma^-(c) \right],
\end{equation}
where the exact formulas for probabilities  $\gamma^+(c)$ and $\gamma^-(c)$ depend on the applied rule. In the following part of the paper, we use the abbreviated notation replacing: $\gamma^+(c)$ by $\gamma^+$, $\gamma^-(c)$ by $\gamma^-$ and $c(t)$ by $c$. Explicit forms of probabilities $\gamma^+,\gamma^-$ are the following:

\begin{enumerate}[(i)]
\item \texttt{GLOBAL\&AND}\\
\begin{eqnarray}
\gamma^+ & = & (1-p)(1-c)c^{2q}+p(1 - c)/2, \nonumber\\
\gamma^- & = & (1-p)c(1-c)^{2q}+p c/2.
\label{eq:gamma_global_and}
\end{eqnarray}

The first component describes conformity, where the change of the state is possible only when two lobbies of size $q$ each (i.e. the total number of agents is equal to $2q$) possess the opposite state than the state of the  voter. The second component is responsible for the change due to the independence.

\item \texttt{GLOBAL\&OR}\\
\begin{eqnarray}
\gamma^+ & = & (1-p)(1-c) \left[ 2\sum_{k=1}^{k=q-1}{{q \choose k}c^{q+k}(1-c)^{q-k}}+ c^{2q} \right] \nonumber \\
& & + \frac{p(1-c)}{2} \nonumber, \\
\gamma^- & = & (1-p)c\left[ 2\sum_{k=1}^{k=q-1}{{q \choose k}(1-c)^{q+k}c^{q-k}} +(1-c)^{2q} \right] \nonumber \\
& & + \frac{p c}{2}.
\end{eqnarray}

Since in the \texttt{OR} case agreement just in one lobby is needed, the sum in the above equations reflects all possible states in which all agents in the $q$-lobby on the one level possess the state opposite to the state of the voter and simultaneously the $q$-lobby on the second level is not homogeneous. \texttt{GLOBAL\&OR} rule for $q=2$ indicates change of the voter's state if three or four of the four agents (two from each level) posses the same state and therefore this it is equivalent to the majority rule \cite{Gal:86,Gal:90,majority_o,majority}.  For  $q > 2$ majority rule is not enough since changes are possible only when at least one lobby is homogeneous, e.g:
\begin{equation}
 \frac{\downarrow \downarrow  \downarrow}{\downarrow  \uparrow \uparrow}  \Uparrow \stackrel{1-p}\longrightarrow    \frac{\downarrow \downarrow  \downarrow}{ \downarrow \uparrow \uparrow} \Downarrow.  
\end{equation}
This fact is direct reason why in the following example there is no change of state: 
\begin{equation}
 \frac{\downarrow \uparrow   \downarrow}{\downarrow  \uparrow \downarrow }  \Uparrow \stackrel{1-p}\longrightarrow  \frac{\downarrow \uparrow   \downarrow}{\downarrow \uparrow \downarrow }   \Uparrow.
\end{equation}

\item 
\texttt{LOCAL\&AND}\\
\begin{eqnarray}
\gamma^+ & = & (1-p)^2(1-c)c^{2q}+p(1-p)(1-c)c^q+\frac{p^2(1-c)}{4}, \nonumber\\
\gamma^- & = & (1-p)^2c(1-c)^{2q}+p(1-p)c(1-c)^q+\frac{p^2c}{4}.
\label{e_and_l}
\end{eqnarray}

Here we have three components: the first describes the situation when the voter behaves like a conformist on both levels, the last one corresponds to the case where on both levels the voter is independent. The second term in Eq.(\ref{e_and_l}) is a mixed one -- the voter behaves as a conformist ($(1-p)c^q$) on one level and simultaneously it is independent on the second level ($p/2$). We multiple this middle term by $2$ since this situation can appear in two configuration: conformist on the first level and independent on the second and vice-versa. 
\end{enumerate}

\section{Results}
Solving analytically rate equation Eq. (\ref{master}) in general (i.e. for arbitrary $q$) is a difficult task. However, it is easy to obtain a numerical solution by iterating Eq.(\ref{master}). In such a way we can obtain the time evolution of the average concentration $<c(t)>$, as well as the stationary value $<c>$. The magnetization $m(t)$ defined by Eq. (\ref{eq_mag}) is directly related to the concentration $c(t)$:
\begin{equation}
m(t)=\frac{N_{\uparrow}(t)-N_{\downarrow}(t)}{N}=2c(t)-1
\label{eq_m_c}
\end{equation}
and therefore using the rate equation (\ref{master}) we can easily find also the average magnetization.
Independently, we can obtain results conducting Monte Carlo simulations and calculating the ensemble average of the magnetization defined by Eq. (\ref{eq_mag1}).

\begin{figure}
\vskip 0.3cm
 \centerline{\epsfig{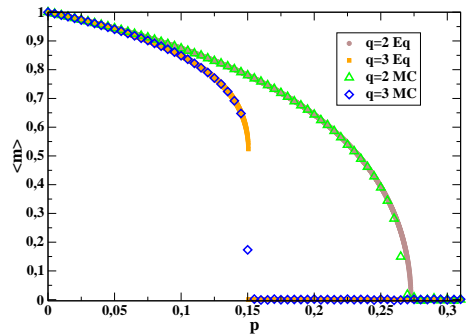}}
    \caption{The average magnetization $<m>$ as a function of the stochastic noise $p$ for the \texttt{GLOBAL\&AND} rule on the duplex clique. Monte Carlo results (empty symbols) were obtained for the system of size $N=10^4$ and averaged over $10^3$ samples. The numerical solutions of Eq. (\ref{master}) are marked with full symbols.}
 \label{and_global1}
\end{figure}

\begin{figure}
\vskip 0.3cm
 \centerline{\epsfig{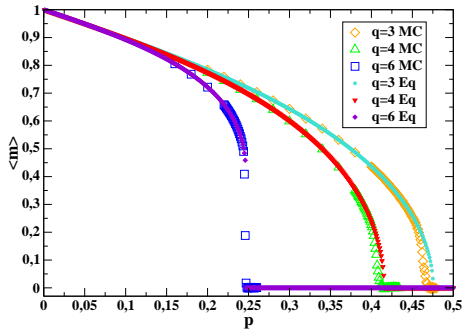}}
    \caption{The average magnetization $<m>$ as a function of the stochastic noise $p$ for the \texttt{GLOBAL\&OR} rule on the duplex clique. Monte Carlo results (empty symbols) were obtained for the system of size $N=10^4$ and averaged over $10^3$ samples. The numerical solutions of Eq. (\ref{master}) are marked with full symbols.}
\label{or_global}
\end{figure}

\begin{figure}
\vskip 0.3cm
 \centerline{\epsfig{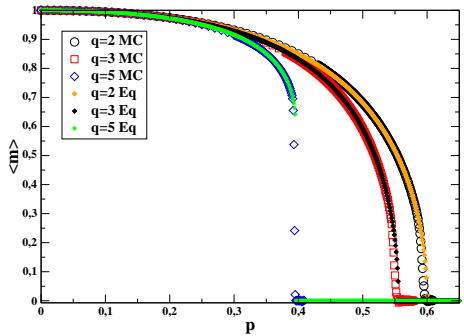}}
    \caption{The average magnetization $<m>$ as a function of the stochastic noise $p$ for the \texttt{LOCAL\&AND} rule on the duplex clique. Monte Carlo results (empty symbols) were obtained for the system of size $N=10^4$ and averaged over $10^3$ samples. The numerical solutions of Eq. (\ref{master}) are marked with full symbols.}
\label{and_local}
\end{figure}

Relations between the average magnetization in the stationary state $<m>$ and the stochastic noise $p$, obtained by two methods (numerical solutions of analytical formulas and Monte Carlo simulations), are presented in Figs. \ref{and_global1}, \ref{and_local} and \ref{or_global}. It is seen that the agreement between the  Monte Carlo results obtained for system size $N=10^4$ and the numerical solution of Eq. (\ref{master}) for the infinite system size ($N \rightarrow \infty$) is very satisfactory.

\section{Phase transitions}
Due to the Landau theory, to describe any kind of phase transition we can introduce the quantity that measure the degree of order (order parameter) \cite{Lan:37,Pli:Ber:98}. Although originally Landau theory was created to describe continuous phase transitions \cite{Lan:37}, it occurred that the theory can be used also in the case of discontinuous phase transitions \cite{Pli:Ber:98}. An order parameter, introduced to distinguish between two phases, is equal to $1$ in the completely ordered state, decreases as a function of the deviation from the order, and becomes zero in the disordered phase. Therefore for our system the natural choice of the order parameter is an average magnetization $<m>$.

It is seen in Figs. \ref{and_global1}, \ref{and_local} and \ref{or_global} that for all three rules \texttt{GLOBAL\&AND}, \texttt{GLOBAL\&OR} and \texttt{LOCAL\&AND}, the system undergoes the phase transition. Below the transition point $p=p_c$, the average magnetization $<m>$ (order parameter) is non-equal zero and above the transition point $<m>=0$. For \texttt{GLOBAL\&AND} (see full symbols in Fig. \ref{and_global1}) the transition changes its type from continuous to discontinuous at $q=3$, which corresponds to $q=6$ for the monoplex clique and thus agrees with results obtained in \cite{Nyc:Szn:Cis:12}.  Analogously, for \texttt{GLOBAL\&OR} (see full symbols in Fig. \ref{or_global}) the transition changes its type also at $q=6$. However, for \texttt{LOCAL\&AND} (see full symbols in Fig. \ref{and_local}), the transition changes its type from continuous to discontinuous already at $q=5$. Moreover, the transition point is much higher than for remaining two rules. The aim of this section is to determine the relation between the threshold value $p_c$ and parameter $q$ for all three rules and understand deeper the nature of the observed phase transitions. 

\begin{figure}
\vskip 0.3cm
 \centerline{\epsfig{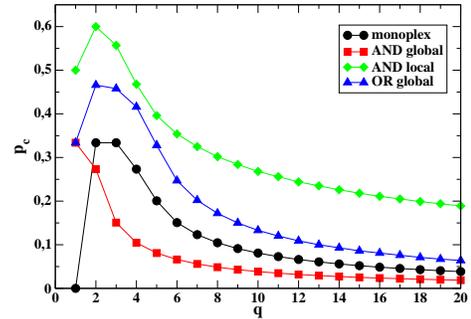}}
    \caption{The relation between the critical value of the stochastic noise $p_c$ and parameter $q$ for different variants of the model.}
        \label{pq}
\end{figure}

We can obtain the the transition point $p=p_c$ directly from the Landau definition of an order parameter $<m>$ using numerical solutions of equation (\ref{master}) or Monte Carlo simulations.  Because, as seen in Figs. \ref{and_global1}, \ref{or_global} and \ref{and_local}, the agreement between Monte Carlo and numerical stationary solutions of Eq. (\ref{master}) is very good, we determine $p_c$ from Eq. (\ref{master}), which is not only much faster but also more accurate method. Figure \ref{pq} shows the relation between the critical value of noise $p_c$ and the size of the $q$-lobby. 

For the monoplex network the critical value of the noise for $q\le 5$ has been derived analytically \cite{Nyc:Szn:Cis:12}:
\begin{eqnarray}
p_c(q) =  \frac{q - 1}{q - 1 + 2^{q-1}},
\end{eqnarray}
which gives $p_c(2)=p_c(3)=1/3$ and $p_c(1)=0$, i.e. there is no phase transition for the linear voter model ($q=1$). However, for the multiplex structure the phase transition is observed even for $q=1$, since we choose one neighbor from each level. Moreover, for $q=1$ the critical value of noise $p$ is equal for \texttt{GLOBAL\&AND} and \texttt{GLOBAL\&OR} rules. This is clear, because in this case, for both rules, the change of the voter's state can happen only if the neighbors chosen from the first and the second level possess the same state, opposite to the voter's state. For \texttt{GLOBAL\&AND} the relation $p_c=p_c(q)$ is a decreasing function for the whole range of $q$, which is also understandable, because this case corresponds to the $q$-voter on monolex with with twice the size of the q-lobby.  

From the point of view of the network structure, certainly the most interesting, among considered rules, is the \texttt{LOCAL\&AND} rule. For both \texttt{GLOBAL} rules multiplex could be in fact replaced by the monoplex network. In the case of \texttt{GLOBAL\&AND}, as already mentioned, we could simply consider the $q$-voter on monoplex with doubled size of the $q$-lobby. The \texttt{GLOBAL\&OR} is less trivial but still could be probably reformulated in terms of the $q$-voter model with the threshold on monoplex \cite{Nyc:Szn:13}. The case of \texttt{LOCAL} independence is not only less trivial, but also the most interesting from the social point of view. It should be remembered that conformity (and simultaneously independence) is relative, i.e. individuals always conform in respect to the particular social group and there are many factors that influence the level of conformity\cite{Mye:13,Cin:Gre:07,Ros:Eri:12,Cia:Gol:04}. It means that the same individual may conform to one group and behave independently in respect to another. For example it has been shown, on the basis of various social experiments, that the level of conformity is much higher in the face-to-face condition than in computer-mediated communication \cite{Cin:Gre:07,Ros:Eri:12}. Hence the idea of local independence is highly justified in modeling social systems. 

\begin{figure}
\centering
\includegraphics[scale=0.45]{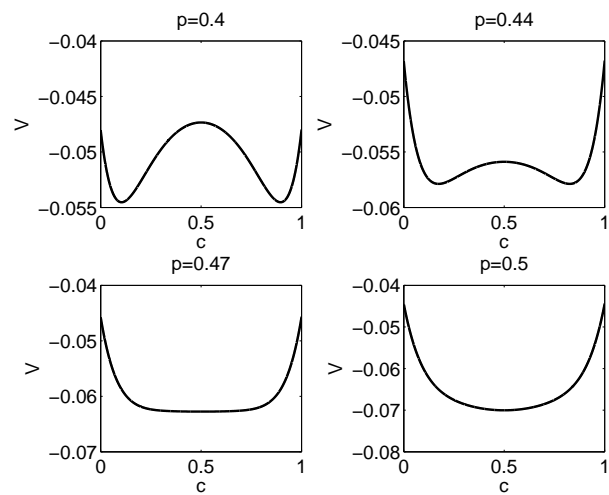} 
\caption{An effective potential $V$, given by Eq. (\ref{pot}), as a function of concentration $c$ for the \textbf{\texttt{LOCAL\&AND} rule and $q=4$}. For small values of noise $p$ potential has $2$ minima that correspond to ordered states (i.e.  $c \ne 1/2$ and simultaneously $m \ne 0$). With increasing $p$, minima are getting shallower and approaching each other. Eventually they form a single minimum, that corresponds to the new disordered phase. This is a typical behavior for a \textbf{continuous phase transition}. \label{pot_q4}}
\end{figure}

\begin{figure}
\centering
\includegraphics[scale=0.45]{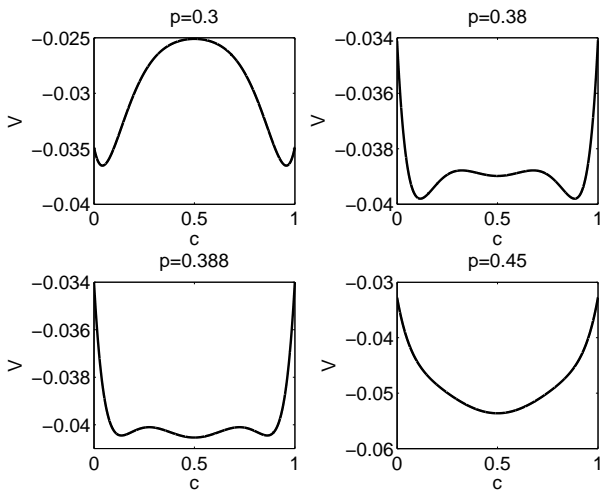} 
\caption{An effective potential $V$, given by Eq. (\ref{pot}), as a function of $c$ for the \textbf{\texttt{LOCAL\&AND} rule and $q=5$}. For small values of noise $p$ potential has $2$ minima that correspond to ordered states (i.e.  $c \ne 1/2$ and simultaneously $m \ne 0$). For larger $p$ the third minimum (that corresponds to the new disordered phase) appears. Initially (for $p \in (p_1^{*},p_2^{*})$) the minima, that correspond to an ordered phase, are deeper than the middle one , i.e. disordered state is metastable. For $p=p_2^{*}$ all three minima are equally deep, i.e. ordered and disordered states are equally probable, which corresponds to the \textbf{discontinuous phase transition}. For $p \in (p_2^{*},p_3^{*})$ the potential has still $3$ minima, but now two ordered states are metastable. Finally, for $p \in (p_3^{*},1)$ the potential has only $1$ minimum that corresponds to the disordered phase. Between spinodal lines i.e. for $p \in (p_1^{*},p_3^{*})$ one can expect hysteresis and indeed it was found in Monte Carlo simulations (see Fig. \ref{pq1}). \label{pot_q5}}
\end{figure}

Therefore, we will now concentrate on \texttt{LOCAL\&AND} rule and discuss the phase transition more thoroughly in this case. First of all let us notice that 
\begin{equation}
F=\gamma^+ -\gamma^-
\label{force}
\end{equation}
can be treated, analogously as in \cite{Nyc:Szn:Cis:12}, as an effective force -- $\gamma^+$ drives the system to the state 'spins up', while $\gamma^-$ to 'spins down'. 
Therefore, inserting explicit forms of $\gamma^+, \gamma^-$ from Eq. (\ref{e_and_l}) we calculate also an effective potential:
\begin{widetext}
\begin{eqnarray}
V & = &  - \int Fdc \nonumber\\
& = & p(1-p) \left\{ q\left[(1-c)c^{q-1} + c(1-c)^{q-1} \right] - \left[c^q - (1-c)^q \right]\right\} \nonumber\\
& + & (1-p)^2 \left\{ 2q\left[(1-c)c^{2q-1} + c(1-c)^{2q-1} \right] - \left[c^{2q} - (1-c)^{2q} \right]\right\} \nonumber\\
& - & p^2/2.
\label{pot} 
\end{eqnarray}
\end{widetext}
To find the critical value of the stochastic noise $p_c$ and the threshold value $\widetilde{q}$, above which the transition becomes discontinuous,  we could now use the Landaus approach, analogously as in \cite{Nyc:Szn:Cis:12}. To do this we first rewrite the potential (\ref{pot}) in terms of magnetization $m$, using relation (\ref{eq_m_c}), and then expand it into power series around $m=0$.  Unfortunately, the form of the potential is much more complex in this case than for the $q$-voter model on monoplex \cite{Nyc:Szn:Cis:12}. Therefore all formulas are much longer and difficult to analyze. To understand the nature of the phase transition, it is much easier and more illustrative to draw potential $V$ as a function of $c$ for different values of $p$ and $q$ (see Figs. \ref{pot_q4} and \ref{pot_q5}).

\begin{figure}
\vskip 0.3cm
\centerline{\epsfig{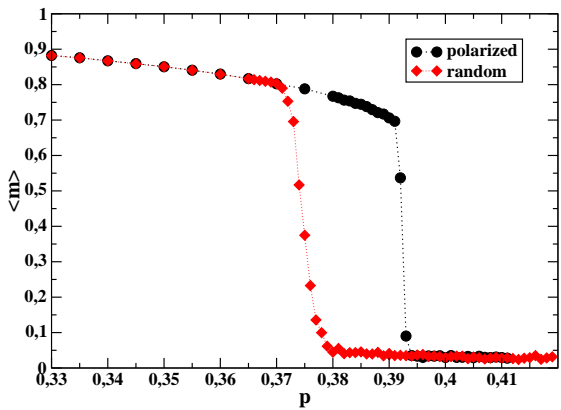}}
\caption{The average magnetization $< m >$ as a function of the stochastic noise $p$ for the \texttt{LOCAL\&AND} rule and $q=5$ obtained from Monte Carlo simulations. Two different initial states were considered --'polarized', i.e. ordered state with $m=1$ and 'random', i.e. disordered state with $m=0$. As expected for discontinuous phase transitions (see Fig. \ref{pot_q5}), hysteresis is observed.}
\label{pq1}
\end{figure}

For $q < \widetilde{q}=5$ (see Fig.\ref{pot_q4}) the potential, given by Eq. (\ref{pot}), behaves typically for the continuous phase transition. Below the transition point the potential has $2$ minima that correspond to ordered states (i.e.  $c \ne 1/2$ and simultaneously $m \ne 0$). With increasing $p$, minima are getting shallower and approaching each other. Eventually they form a single minimum, that corresponds to the new disordered phase. For $q \ge \widetilde{q}$ the potential indicates discontinuous phase transition. For small values of noise $p$ potential has $2$ minima that correspond to ordered states (i.e.  $c \ne 1/2$ and simultaneously $m \ne 0$). For larger $p$ the third minimum (that corresponds to the new disordered phase) appears. Initially (for $p \in (p_1^{*},p_2^{*})$) the minima that correspond to an ordered phase are deeper than the middle one , i.e. disordered state is metastable. For $p=p_2^{*}$ all three minima are equally deep, i.e. ordered and disordered states are equally probable, which corresponds to the discontinuous phase transition. For $p \in (p_2^{*},p_3^{*})$ the potential has still $3$ minima, but now two ordered states are metastable. Finally, for $p \in (p_3^{*},1)$ the potential has only $1$ minimum that corresponds to the disordered phase. Between spinodal lines i.e. for $p \in (p_1^{*},p_3^{*})$ one can expect hysteresis and indeed it was found in Monte Carlo simulations (see Fig. \ref{pq1}).

\section{CONCLUSIONS}

We have generalized the $q$-voter model with independence (noise) $p$ to a duplex clique, i.e. a network that consists of two distinct levels (layers), each of which is represented by a complete graph (i.e. a clique) of size $N$. Levels represent two different communities (e.g. Facebook and school class), but are composed of exactly the same people -- each node possesses a counterpart node in the second level. Such an assumption reflects the fact that we consider fully overlapping levels, being an idealistic scenario. We also assume that each node possesses the same state on each level, which means that the society consists of non-hypocritical individuals only.
We have considered two criteria of level dependence -- one related to the status of independence (GLOBAL vs. LOCAL) and one related to peer pressure (AND vs. OR). The \texttt{GLOBAL} rule means that an agent is independent on both levels, but the \texttt{LOCAL} rule admits a situation where a person is independent in one clique but not in the other. Furthermore, the \texttt{AND} dynamics is more restrictive and a node changes its state only if {\it both levels} suggest changes, in the \texttt{OR} variant {\it one level} is enough to change the state of an individual.  

For all three considered rules (\texttt{GLOBAL\&AND}, \texttt{GLOBAL\&OR} and \texttt{LOCAL\&AND}), the system undergoes a continuous order-disorder phase transition at $p=p_c(q)$ for $q<\widetilde{q}$ and a discontinuous for $q \ge \widetilde{q}$, where $p_c$ and $\widetilde{q}$ are rule-dependent. The \texttt{GLOBAL\&AND} rule leads to a trivial result, identical with the monoplex case for a doubled value of $q$. For the \texttt{GLOBAL\&OR} dynamics, $p_c$ is larger than for the monoplex network. However,  $\widetilde{q}$ is identical with the monoplex case, i.e. $\widetilde{q}=6$. In contrast to the other two rules, we find a qualitative change for the \texttt{LOCAL\&AND} rule, as the phase transition becomes discontinuous for $\widetilde{q}=5$. The case of \texttt{LOCAL} independence is not only less trivial, but also more interesting and better justified from the social point of view.
In particular, it has been shown that the level of conformity during face-to-face communication is significantly higher than during computer-mediated communication such as the Internet \cite{Cin:Gre:07,Ros:Eri:12}. 

This suggests that the \texttt{LOCAL\&AND} rule is the most suitable for real social systems. Certainly it could be further developed by introducing different values of noise on each level. The simplistic duplex clique topology, as introduced in this paper, can be also modified to obtain a more general network. For instance, one could  consider partially overlapping cliques, where some nodes possess no counter-node on the second level. 
Unfortunately, these modifications significantly complicate the model by introducing additional parameters and therefore are beyond the scope of this paper. However, even considering such a simple model as here, we can observe that a multiplex network can introduce significant differences in opinion dynamics.

\begin{acknowledgments}
This work was supported by funds from the National Science Centre (NCN, Poland) through post-doctoral fellowship no. 2014/12/S/ST3/00326 (to AC) and grant no. 2013/11/B/HS4/01061 (to KSW).
\end{acknowledgments}

\end{document}